\documentclass[aps,pra,twocolumn,superscriptaddress,reprint]{revtex4-1}

\usepackage[T1]{fontenc}
\usepackage{ae,aecompl}
\usepackage[english]{babel}
\usepackage{subfigure,bm,dcolumn,graphicx,hyperref}
\usepackage{amsmath}
\usepackage{epstopdf}
\usepackage{color}
\usepackage[normalem]{ulem}

\renewcommand{\vec}[1]{\mathbf{#1}}

\begin{document}
\title{Inducing and controlling magnetism in the honeycomb lattice 
through harmonic trapping potential}

\author{K.~Baumann}
\affiliation{Scuola Internazionale Superiore di Studi Avanzati (SISSA), 
  Via Bonomea 265, 34136 Trieste, Italy}

\author{A.~Valli}
\affiliation{Scuola Internazionale Superiore di Studi Avanzati (SISSA), 
  Via Bonomea 265, 34136 Trieste, Italy}
\affiliation{CNR-IOM DEMOCRITOS, Istituto Officina dei Materiali, Consiglio Nazionale delle Ricerche}
\affiliation{Institute for Solid State Physics, Vienna University of Technology, 1040 Vienna, Austria}

\author{A.~Amaricci}
\affiliation{Scuola Internazionale Superiore di Studi Avanzati (SISSA), 
  Via Bonomea 265, 34136 Trieste, Italy}
\affiliation{CNR-IOM DEMOCRITOS, Istituto Officina dei Materiali, Consiglio Nazionale delle Ricerche}

\author{M.~Capone}
\affiliation{Scuola Internazionale Superiore di Studi Avanzati (SISSA), 
  Via Bonomea 265, 34136 Trieste, Italy}
\affiliation{CNR-IOM DEMOCRITOS, Istituto Officina dei Materiali, Consiglio Nazionale delle Ricerche}


\date{\today}
\begin{abstract}
We study strongly interacting ultracold spin-$1/2$ fermions in a honeycomb lattice 
in the presence of a harmonic trap. 
Tuning the strength of the harmonic trap we show that it is possible
to confine the fermions in artificial structures reminiscent of graphene nanoflakes in solid state. 
The confinement on small structures induces magnetic effects 
which are absent in a large graphene sheet.
Increasing the strength of the harmonic potential we are able to induce different magnetic states, 
such as a N\'eel-like antiferromagnetic or ferromagnetic states, 
as well as mixtures of these basic states. 
The realization of different magnetic patterns is associated with the
terminations of the artificial structures, in turn controlled by the confining potential.
\end{abstract}

\maketitle

\section{introduction}\label{intro}

The discovery of graphene~\cite{geimSci324,novoselovNat438}, 
a two-dimensional sheet of carbon atoms arranged on a honeycomb lattice, 
has generated a new field of research which attracted an unprecedented 
interest as it combines an almost ideal non-trivial quantum problem 
with extraordinary mechanical, electronic, thermal, 
and transport properties~\cite{novoselovNat438,castro_netoRMP81}. 
From a more theoretical perspective, the rise of graphene has increased the interest in the properties
of quantum particles on the honeycomb lattice. 

Besides the synthesis of a variety of graphene-based systems, this has also pushed the community 
to devise quantum simulators of the honerycomb lattice using ultracold atoms~\cite{leePRA80,tarruellNat483,ducaSci288,Soltan-Panahi},
which can access regimes which are not easily reached in solid state systems.

One of the most elusive properties of  solid-state graphene systems is magnetism. 
Infinite, or very large, graphene sheets do not show magnetic ordering~\cite{chenSR3}, which 
is instead proposed and realized in {\it nanoscopic} structures composed by a small number of carbon atoms
when the termination have a so-called zigzag pattern~\cite{nakadaPRB54,rossierPRL99}. However, the instability of zig-zag edges severely
limits the realization of magnetic graphene nanostructures and the first solid experimental evidence is very recent~\cite{slotaNat557} .

Promising candidates for magnetism are nanoflakes~\cite{valliPRB94,koopPRB96}. 
Their theoretical phase diagram is quite rich, and it is characterized by a strong competition between short-range antiferromagnetic (AF) 
and long-range ferromagnetic (FM) correlations. The former are particularly strong in insulating half-filled flakes with one fermion per site, 
while the latter emerge when the density is reduced and the carriers become more mobile.
In principle this competition might be exploited to  manipulate the magnetic ground state by,  e.g., electrostatic~\cite{kabirPRB90,valliPRB94,gangulyPRB95} or chemical doping~\cite{junAPL98,zouSR5,ortizPRB94} and to engineer different kinds of spin filters~\cite{junAPL98,caoAPL99,zouSR5,lundebergNatPhys5,valliNL18,valli1810.11307}. This rich scenario is however so far largely unexplored owing to the technical difficulties to control the edges of solid-state graphene nanosystems.

%~\cite{chenSR3,magdaNat514,tucekNatComm8}
%seem to support the theoretical predictions, 
%solid evidence for edge magnetism has been reported only very
%recently in nanoscopic graphene ribbons in a solution~\cite{slotaNat557}.

\begin{figure}[htbp!]
\centering
  \includegraphics[width=1\linewidth, angle =360]{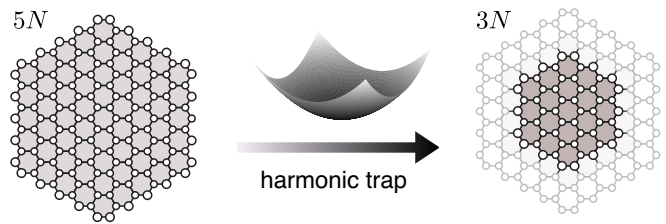}
 \caption{(Color online) Protocol for the realization of artificial nanoflakes 
 by spatial confinement using a optical trapping potential. 
 Solid and shaded lines denote the $5N$ and $3N$ hexagonal flakes, respectively.}
 \label{fig:1}
\end{figure}

In this paper we propose  an alternative route to induce
magnetism in effective artificial graphene
structures formed by cold atoms moving in optical lattices~\cite{blochRMP80,mazurenkoNat545} which overcomes
the limitations of solid-state realizations.
The idea is to engineer an optical lattice with the graphene honeycomb
structure~\cite{leePRA80,tarruellNat483,ducaSci288,Soltan-Panahi}  
in the presence of a strong harmonic  trapping potential which confines the
fermions in a limited portion of the lattice, thus realizing an 
artificial {\it nanostructure}. A schematic illustration of this idea
is depicted in Fig.~\ref{fig:1}. 
In what follows we show that this procedure leads to sufficiently well
defined artificial edges which mirror the different edges of solid-state nanosystems. 
We demonstrate that, by continuously tuning the strength of the trapping
potential, and thus the size of the artificial flake, we can induce magnetic phases starting from a non-magnetic system
and we can induce transitions between different magnetic states
with different arrangements of the spins. We will highlight that the effective flakes will be characterized by 
a spatially  inhomogeneous distribution of the fermions, with a larger local density in the central region. This will
enrich the scenario of the magnetic properties of the effective flakes with respect to  graphene nanoflakes.

The paper is organized as follows: In Sec. \ref{m_and_m} we introduce the model used to describe the graphene-like structure and give an overview of dynamical mean-field theory that is used to solve the system.  In the Sec. \ref{results} we discuss the two main results, namely the creation of the artificial nanostructures in optical lattices reminiscent of graphene nanoflakes (Sec. \ref{art_nano}) and the magnetic properties of the synthetic nanoflakes (Sec.\ref{mag}). Sec. IV is dedicated to concluding remarks.

%The simple and controlled experimental framework of ultracold atoms would 
%allow to directly test these predictions providing important information for solid-state
%material design.
\section{model and method}\label{m_and_m}

We model our artificial structure by a Fermi-Hubbard model with
repulsive interactions on a two-dimensional honeycomb lattice 
\begin{equation}
\begin{split}
  \mathcal{H} = &-t \sum \limits_{\langle i,j \rangle,\sigma} (\hat{c}^{\dagger}_{i\sigma}\hat{c}_{j\sigma}+h.c) + U\sum \limits_{i}\hat{n}_{i\uparrow}\hat{n}_{i\downarrow} \\
  &+ \sum \limits_{i} (V_i-\mu) (\hat{n}_{i\uparrow}+\hat{n}_{i\downarrow}),
\end{split}
\end{equation}
where $\hat{c}^{\dagger}_{i\sigma}$ ($\hat{c}_{i\sigma}$) denotes the
creation (annihilation) operator for spin-1/2 fermions  and $\hat{n}_{i\sigma}$ is
the density operator at site $i$ for the two spin states, 
labeled by $\sigma \in \{ \uparrow, \downarrow \}$. $n_i = \sum_{\sigma} n_{i\sigma}$ is the total density on site $i$.
$t$ denotes the nearest-neighbour tunneling amplitude,  
$U$ the on-site repulsion while the chemical potential $\mu$ controls the number fermions.  
$V_i = V_0 r_i^2$ is a harmonic trapping potential,  with $\vec{r}_i = (x_i, y_i)$ the lattice coordinate and $r_i = \vert \vec{r}_i\vert$. In our calculations we consider a $L=150$ sites lattice with hexagonal
shape and zigzag edges whose geometric center, $\vec{r}=(0,0)$,  
coincides with the minimum of the parabolic potential, as shown in the left panel of Fig.~\ref{fig:1}.

In the absence of the optical trap, on an infinite honeycomb lattice, 
this model displays a transition from a Dirac semimetal to an antiferromagnetic insulator 
at average density of one fermion per site (half-filling)~\cite{sorellaEPL8}. 
A metallic state is generally found for any other number of fermions. 
The  critical interaction strength for the onset of magnetism has been
estimated to be $U_c \simeq 3.87 t$~\cite{yunokiPRB74} via numerically
exact Quantum Monte Carlo simulations of large lattices. 
On smaller systems with zigzag edges, an AF spin
ordering establishes at the boundaries~\cite{rossierPRL99,kabirPRB90,valliPRB94}. 
Importantly, theoretical~\cite{valliPRB94} and experimental~\cite{magdaNat514} 
evidences suggest that this edge magnetism survives up to room temperature. 
As expected, AF order is favoured in half-filled systems, for which
the Hubbard interaction is more effective in localizing the carriers. The general expectation is that, when the density deviates
substantially from half-filling, delocalized metallic states are favoured. The free motion of carriers
in a metallic state destroys the AF ordering, while they it can coexist with a FM order, which is not
spoiled by the motion of carriers. This leads to a competition between AF and FM tendencies which is mainly determined
by the density. It is therefore very interesting to address this issue in our inhomogeneous system where the local density
changes as we move from the center to the edge of the system.

We solve the model using a real-space dynamical mean-field theory
(DMFT)~\cite{georgesRMP68,caponePRB76,weberPRB86,snoekNJP10} approach, 
which has been previously used to study inhomogeneous systems, 
such as cold atoms~\cite{snoekNJP10,amaricciPRA89,gorelikPRL105}, 
and nanostructures~\cite{valliPRL104,valliPRB86,schuelerEPJST226}, 
including isolated graphene nanoflakes~\cite{valliPRB94}. 
In the homogeneous case DMFT approximates the lattice self-energy 
of the interacting many-body problem with a local, momentum-independent, 
self-energy which, however, retains the full frequency dependence which allows to 
capture non-trivial quantum correlations characteristic of strongly correlated systems~\cite{georgesRMP68}.
In order to treat an intrinsically inhomogeneous system and the confinement effects
induced by the parabolic potential we need to relax  this approximation using real-space 
DMFT, in which the self-energy remains local but it acquires a dependence on the specific lattice, 
i.e., $\Sigma_{ij\sigma} = \delta_{ij}\Sigma_{i\sigma}$. 
%This allows to describe the different behavior of bulk and edge sites 
%as well as the emergence of magnetic exchange correlations 
%following the electronic charge redistribution 
%due to the trap and the Coulomb repulsion. 

In order to explore extensively the dependence of parameters we focus
on finite artificial flakes, following previous calculations in a
solid-state set-up~\cite{valliPRB94,valliNL18}. We borrow from these works
the choice to start from the 150-site cluster that we label as $5N$ (according to a notation where $N$ is the number of sites on an edge). This
cluster  contains smaller hexagonal nanoflakes ($4N$, $3N$, ...) with the same symmetry, 
as well as flakes with different edge termination (bearded), 
which are shown in Fig.~\ref{fig:2}(f), which can all in principle be stabilized by the trapping potential.

This allows to prove whether our protocol to confine atoms in effective nanostructures actually works in 
configurations which have already been tested. In this light
we refrain from a detailed comparison with potential realizations with actual cold-atom experiments,
which we postpone to future research. We notice that a similar scheme, where the optical trapping is used 
to induce quantum states in an optical lattice loaded with ultracold fermions, has been realized in Ref. \cite{Chiu}.
%to a future work where we will address larger lattices and specific choices 
%of the nature and number of trapped atoms, the trapping potential and the optical lattice.  

\section{Results}\label{results}

\subsection{Artifical Nanostructures}\label{art_nano}
As discussed in the following, the electronic distribution and the magnetic ordering 
of the fermions in the trapping potential depend on the value of the local repulsion, which 
in cold-atom experiments can easily be controlled via tuning the strength of the optical lattice
and the scattering length, when Feshbach resonances are available.
We consider two case: $U/t = 3.75$ and $U/t = 11.25$. 
The first value corresponds to a realistic choice for actual graphene, 
and it falls in the range where the semimetal/AF insulator transition takes
place for the infinite honeycomb lattice~\cite{yunokiPRB74,sorellaEPL8}.
The latter is sufficient to put any graphene structure deeply in the
Mott state, where the fermions are described as localized spins
interacting via a Heisenberg exchange.

\begin{figure}[]
\centering
  \includegraphics[width=\linewidth, angle =360]{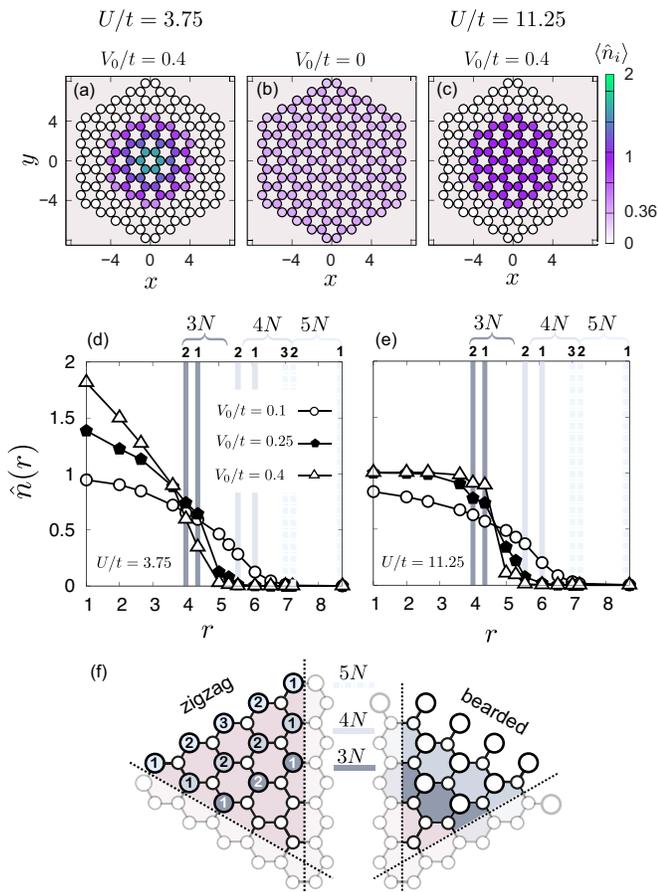}
 \caption{(Color online) Spatial confinement into artificial nanoflakes. 
 (a-c) Map of the local density $n(\vec{r}_i) = \langle n_i \rangle$ 
 in the absence of the trap,  and for trapping potential strengths $V_0/t=0.4$ 
 at repulsion $U/t = 3.75$ and $U/t = 11.25$. 
 (e-d) Radial density distribution for selected potential strengths 
 $V_0/t = 0.1, 0.25, 0.4$, at $U/t = 3.75$ and $U/t = 11.25$. 
 Average fermion density: $\langle n \rangle = 0.36$. 
 The vertical solid lines in panels (d-e) mark the position of the edge sites, 
 as labelled in panel (f), where only one sixth of the flake is shown.  }	
  \label{fig:2}
\end{figure}

We will first consider a system with the same number of fermions per each spin species $N_{\uparrow} = N_{\downarrow} = N_{f}/2$, where $N_f$ is the total number of fermions, which in fixed. This is the standard situation for a cold-atom experiment, where the 
number of fermions in each species is conserved. As a second step we will also release this constraint, allowing the system to relax in a state
with a finite global magnetization (i.e., with $N_{\uparrow} \neq N_{\downarrow}$). This situation would be realized in systems in which spin-flip processes are possible as a consequence of the coupling to an external environment or the inclusion of a small artificial spin-orbit coupling. From a theoretical perspective, this will be useful to clarify the tendency towards ferromagnetic ordering with a finite magnetization for some values of the interaction and of the trapping potential.

In all our calculations we consider $N_{f} = 54$, which coincides with the number of sites composing the $3N$ nanoflake with zigzag edges. 
Thus, if the fermions can be trapped in the portion of space
corresponding to the $3N$ flake we would have one fermion per site (half-filling), 
which is the ideal situation for the onset of AF ordering, at least for a homogeneous system.  
On the other hand, for our system of 150 sites, the average filling is $n  = N_{f}/L = 0.36$, i.e. a small
density which makes interactions marginally effective. Therefore, in the absence of trapping potential we have no 
magnetic effects and the fermions are spread over the whole system also for large interaction strength.

This non-magnetic state is the starting point to introduce the harmonic potential. 
Increasing the strength $V_0$, we progressively localize the fermions in the central region. 
This is demonstrated in Figs.~\ref{fig:2}(a-c), where we show the map 
of the local density $\langle n_i \rangle$ on the whole system for calculations with $N_{\uparrow} = N_{\downarrow}$\footnote{When we release the equal-population constraint the changes in the density profile are very small and they are basically invisible in the plots}.
In the absence of the trap we recover a nearly homogeneous system (the small deviation is due to boundary effects) with $\langle n_i \rangle \approx 0.36$ for every site $i$. Conversely for trapping potential strength $V_0/t=0.4$ the fermions 
are spatially localized within a reduced region around the center of the trap. 
The sharpness of the  confinement depends on the value of the Coulomb repulsion. 

In order to investigate this aspect, 
in Figs.~\ref{fig:2}(d,e) we show the radial profile of the local density 
as a function of the distance from the trap center $r=|\vec{r}|$ 
for different values of the trapping potential.  
In shallow traps (e.g., $V_0/t=0.1$ and $V_0/t=0.25$) 
the fermion distribution is only weakly affected 
by the Coulomb repulsion due to the low average local density. 
As the trap deepens, the fermions tend to leave the boundaries 
to pack in the central region, 
and we witness an important effect of the repulsive interaction, 
which competes with the spatial charge accumulation. 
This effect is clear at $V_0/t =0.4$. 
For relatively weak repulsion $U/t = 3.75$, as in Figs.~\ref{fig:2}(a,d),  
the charge accumulates towards the center of the trap, 
where we approach the maximum local density allowed by Pauli principle.
Therefore, even if the fermions are localized in a region $r \lesssim 5$, 
the distribution is not uniform within the confinement region. 
At $U/t = 11.25$ instead, as in Figs.~\ref{fig:2}(c,e), 
the fermions are Mott localized, and energetically costly double occupancies 
$\langle n_{i\uparrow} n_{i\downarrow} \rangle$ 
are strongly suppressed throughout the lattice. The tendency to reduce double 
occupancy contrasts the packing of fermions in the center of the trap and favors
single occupation $\langle n_i \rangle =1$. 
In the strong-coupling (Mott) regime, 
most of the fermions are confined within a region $r \lesssim 4.5$, 
characterized by a \emph{nearly constant} local density of $n_i \simeq 1$,  
with a sharp drop at this effective boundaries. 
The competition between the trapping potential and the repulsion
results in a smaller structure, almost homogeneously filled, 
which appears as a promising effective nanoflake. 
Our results directly reflect the incompressible nature of the Mott
insulator, where the configuration with singly-occupied sites is
highly favoured and robust with respect to the packing effect of the
trapping potential. In contrast, in the weak- and intermediate-coupling regimes 
the system is a quantum fluid with a finite compressibility, 
which allows charges to accumulate in the center of the trap.

\begin{figure}[htbp!]
\centering
  \includegraphics[width=0.9\linewidth, angle =360]{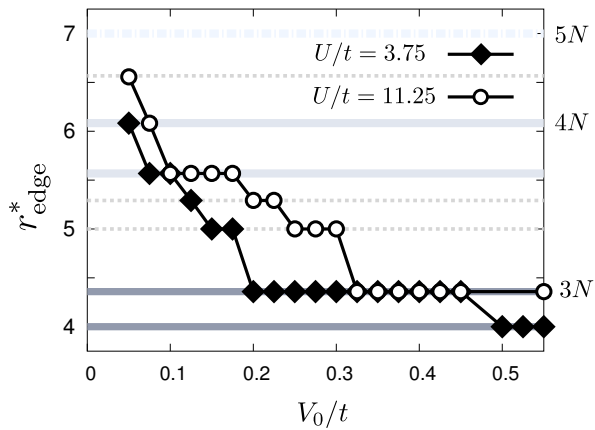}
\caption{(Color online) Effective edge $r^{*}_{\textrm{edge}}$ 
  of the arificial flake induced by the trapping potential as a function of $V_0/t$. 
  Horizontal lines correspond to the positions of the edge sites as labelled in Fig. 2(b). 
  In this setup, the $3N$ flake is stable for a wide range of $V_0/t$. } 
\label{fig:3}
\end{figure}

In Figs.~\ref{fig:2}(d,e) we denoted by vertical lines the radii 
corresponding to the edge sites of different hexagonal nanoflakes 
($5N$, $4N$, and $3N$) as labeled in Fig.~\ref{fig:2}(f).  
At $V_0/t =0.4$, the sharp edge observed in the strong coupling regime 
coincides with that of the 3N nanoflake with zigzag edges which we already 
identified as  a promising candidate for the emergence of magnetism because
it can a host a half-filled configuration for the chosen total number of particles. 

Before probing the magnetic properties of the artificial flake, 
we introduce a concrete definition of the effective edge $r^*_{\textrm{edge}}$. 
In Fig.~\ref{fig:3} we plot as a function of $V_0/t$ the estimate of $r^*_{\textrm{edge}}$ given by the position
where the derivative of the density with respect to the position $\partial n/\partial r$ is maximal. 
Obviously, this definition has a degree of arbitrariness, but we have verified that different criteria provide the same
result for large interactions, while some intrinsic ambiguity is present for small interactions.
%We have verified that the definition of the effective boundary does
%not depend substantially on the chosen criterion~\footnote{For strong repulsion 
%the estimate obtained with this definition coincides
%with alternative options, such as the position for which the
%derivative of the density with respect to the potential.
%$\partial n/ \partial V_0$ is maximum. 
%Some discrepancy is instead found for weak repulsion  
%as detailed in the Supplemental Material.}. 
%I would not add this one...
%[...] and the position where the density
%becomes smaller than an arbitrary threshold  $n(\vec{r}_i) < 0.25$. }.
We marked as horizontal lines the positions of the edge sites as in Fig.~\ref{fig:2}. 
Upon increasing $V_0/t$ the fermionic cloud is attracted 
towards the center of the trap and its effective size is reduced. 
The contraction is faster at weak coupling 
with respect to strong coupling because the Coulomb repulsion 
acts as internal pressure competing with the trapping potential. 
Interestingly, in both regimes the system evolves 
through a series of effective flakes of different sizes. 
The 3N zigzag-edged flake is the most stable artificial structure, 
due to the initial choice of $N_f$, 
and it is realized in a wide range of trapping potential strength.

\subsection{Magnetism}\label{mag}
\begin{figure}[]
\centering
  \includegraphics[width=1.0\linewidth, angle =360]{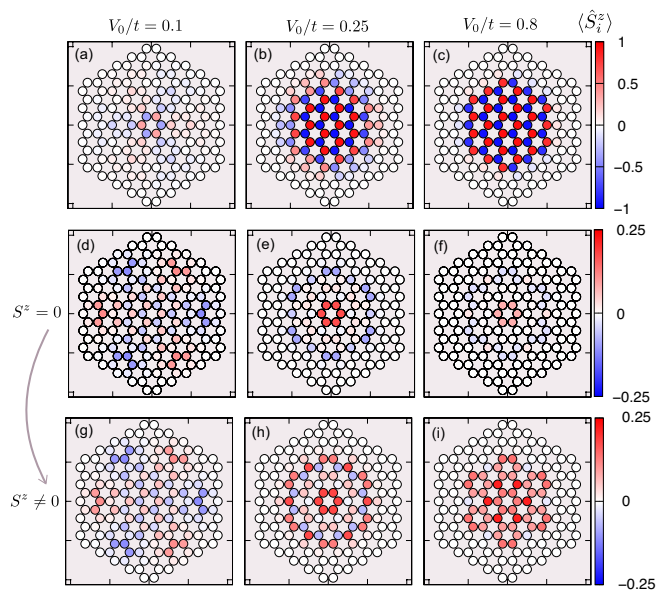}
 \caption{(Color online) Map of the local magnetization $\langle S^z_i \rangle$. The first row (a-c) presents results for $U/t=11.25$, 
 while the second row (d-f) shows data for $U/t=3.75$ with $\langle S_z \rangle =0$ (d-f)  and the third row (g-i) data for  $U/t=11.25$  with $\langle S_z \rangle \neq 0$. In all cases we show results for three different values of $V_0/t$.
 The competition of emergent AF and FM magnetic exchanges 
 gives rise to different magnetic states, depending on the electronic density distribution 
 determined by $V_0/t$ and $U/t$. }
 \label{fig:4}
\end{figure}

We are now in the position to test whether the effective nanostructures that we have defined actually support magnetic ordering. Therefore in Fig.~\ref{fig:4} we report color maps of the local magnetization along the $z$-direction 
$\langle S^z_i \rangle = \langle n_{i\uparrow} - n_{i\downarrow} \rangle$ for the two values of interaction and three values of trapping potential used in Fig. \ref{fig:1}. Red (blue) indicates a positive (negative) magnetization. 
Since the honeycomb lattice is bipartite, i.e., it can be divided in two sub-lattices A and B, such that sites of A are only connected with sites of B and vice versa, we can define a perfect AF state when the magnetization of sublattice A is the opposite of that of sublattice B $\langle S^z_{i_A}\rangle = - \langle S^z_{i_B}\rangle$, while a FM has the same magnetization on the two sublattices. A ferrimagnetic state is characterized by the presence of both a staggered and a uniform magnetization.  In a non-magnetic state we have vanishing local magnetic moments 
$\langle S^z_{i_{A,B}}\rangle \approx 0$. 
The results in Fig.~\ref{fig:4} show a rich landscape of magnetic
solutions which depend from the effective radius and the density distribution 
of the artificial flakes.

%\begin{figure*}[]
%\centering
%  \includegraphics[width=0.8\linewidth, angle =360]{Fig4.eps}
% \caption{(Color online) Magnetization along the z direction $\langle
%   S^z_i \rangle$ for selected potential strengths and two interaction
%   regimes $U/t = 3.75$ and $U/t = 11.25$. Red (blue) marks positive
%   (negative) magnetization. A series of magnetic states
%   (paramagnetic, antiferromagnetic, ferromagtic/ferrimagnetic, mixed)
%   are shown. Parameters: $V_0/t = 0.100; 0.125; 0.400$ from left to
%   right for a fixed filling of $\langle n_i \rangle = 0.36$.}
% \label{fig:4}
%\end{figure*}

We start presenting results where we impose that the total number of up and down fermions is conserved and $N_{\uparrow} = N_{\downarrow} = N_f/2$, or $\langle S^z \rangle = 0$, where $S^z = \sum_i S^z_i$. 
We first consider results in the strong-coupling regime, where the emergence of an effective half-filled 3N nanoflake is clear, 
as shown by our data for $U/t = 11.25$.  
In the absence of a trapping potential, or for very shallow ones, the fermion density is low on every site 
(we remind that $\langle n_i \rangle \approx 0.36$ at $V_0=0$) 
and the flake is not magnetic (not shown). 
As the fermions are attracted towards the center of the trap, 
as in Fig.~\ref{fig:4}(a,b), we observe a clear AF pattern 
in the region of the inner rings, where $\langle n_i \rangle \approx 1$ 
while in the region where $\langle n_i \rangle \approx 0.5$ (quarter-filling)
we find weakly FM islands. 
In Fig.\ref{fig:4}(c) we show that increasing $V_0/t$ 
leads to a dramatic enhancement of the magnetic moments, 
which form N\'eel AF state extended over the whole occupied central region. 
The character of the magnetic ordering of this half-filled structure is not surprising and it 
can be easily understood in
terms of the strong-coupling limit $(U/t \gg 1)$ of the Hubbard model, 
that leads to an effective Heisenberg model with a nearest-neighbor exchange coupling $J=4t^2/U$
which leads to an AF ordering on the honeycomb lattice. 
The magnetic response confirms the formation, in the strong-coupling regime,  
of a well-defined effective flake, with properties reminiscent of an isolated graphene nanoflake, 
and proves that a simple change of the trapping potential can induce a magnetic state. The results in the 
strong coupling limit do not change if we relax the $\langle S_z\rangle = 0$ constraint, confirming that the
AF solution is the actual groundstate of the system.

At intermediate interaction (e.g., for $U/t = 3.75$) the evolution of the magnetic
properties in Fig~\ref{fig:4}(d-f) is richer and more peculiar, as a result of a less pronounced tendency 
towards the formation of magnetic moments and the more ambiguous definition of an artificial edges. 
On the other hand,
as discussed in previous work on graphene nanoflakes\cite{valliPRB94}, densities different from one fermion per site favor FM correlations
which are more compatible with metallic behavior. In our artificial nanoflake, this leads to a non-trivial evolution as a function of $V_0/t$ characterized
by the competition between AF and FM correlations.

For shallow traps, the fermions are spread over a relatively large area, and AF ordering establishes in the central region,  
even if weaker than in the large-$U$ case. In the outer region we observe the development of small FM islands which 
are ordered in a staggered pattern around a hexagonal ring as shown in Fig.~\ref{fig:4}(d). 
This suggests the emergence of long-range AF correlations between FM domains 
at distances $d_{AF}\approx 5r$,  
mediated by the short-range (i.e., $d_{AF} = r$) Heisenberg exchange. 
For deeper traps the packing towards the center is stronger because of the
smaller repulsion. 
Once the occupation of the sites in the inner region becomes close to 2 the tendency towards AF order is washed away, and the center of the trap tends to host a FM island surrounded by a nonmagnetic region and an external ring where the fermions have the opposite spin with respect to the center. 
When the trapping potential becomes even larger  (Fig.~\ref{fig:4}(f) for $V_0/t=0.8$) the magnetic moments become smaller
because of the increased density in the center of  the trap.

Our results directly demonstrate that tuning the trapping potential 
and the interaction strength we can induce magnetism or change completely its nature.
For example if we change the interaction strength at fixed $V_0/t$  we can induce a
transition between a FM and AF states (panels (b) and (e) or (c) and (f) of Fig.~\ref{fig:4}).

Finally, as anticipated above, we relax the the $\langle S_z\rangle  = 0$ constraint. This protocol,
which would require to include some mechanism able to flip the spins of the atoms, is chosen in 
order to better highlight the tendency towards FM that we have identified for the case with moderate
electron-electron interaction. In contrast with the large-$U$ case, here we find that the system indeed
minimizes its energy by polarizing the fermionic spins. In the third row of Fig.~\ref{fig:4} we show these
results. In particular if we compare panel (h) with panel (e) and panel (i) with panel (f) we find a clear 
enhancement of the FM correlations which invade a larger portion of the effective flake and, for the
largest value of the potential that we considered we obtain a FM polarization spread over the whole
effective flake. 

%This nanoflake is found to be in a mixed configuration between 
%a ferro- and a ferrimagnetic state, where the zigzag edges exhibit short-range AF order, 
%but the flake is characterized by a net uncompensated magnetic moment. 

%Importantly, by comparing the results from Figs.~\ref{fig:4}(c,f) 
%we demonstrate that, by trapping the fermions in an artificial flake, 
%it would theoretically be possible to tune the magnetic state between AF and FM 
%just by tuning the interaction by Feshbach resonance. 

%\cc{***}The evolution of the magnetism in the
%weak-interaction regime, i.e. evolving from the edges to the entire
%effective nanoflake, confirms the relevance \cc{(of the nature)} of the edges
%in the development of magnetic ordering. 
%\cc{I would probably remove this...}

\section{Conclusions}
In this work we have shown that a trapping potential can induce a
variety of magnetic phases in an otherwise non-magnetic honeycomb lattice. 
In particular, a parabolic potential can be used to trap the fermions 
in artificial nanoflakes, which inherits the properties 
that have been widely studied in a solid-state framework. 
The trapping is most effective for strong fermion-fermion repulsion, 
underlining the important effect of interactions in the realization 
of well-defined artificial edges. 
Our work shows a novel route to induce magnetism in artificial graphene nanostructures, 
and it is expected to be robust with respect to details of the system, 
i.e., actual size and number of fermions as long as the interactions 
can be made sufficiently strong to reach the Mott regime.  

Also when it is difficult to establish a direct correspondence with solid-state systems, 
because the trapping leads to strongly inhomogeneous density distributions, 
we find a competition between FM and AF tendencies. 
This leads to a tunable system evolving from a weak antiferromagnet to a ferromagnet. 
The magnetic ordering is also expected to be reflected in the transport properties,
leading to highly non-trivial spin transport.
The possibility to induce different magnetic states 
could be exploited to investigate of spin-filter~\cite{valliNL18} and spin-valve effects
within transport experiment in optical lattices~\cite{greinerPRA63,mandelPRL91}.

We  acknowledge support from H2020 Framework Programme, 
under ERC Advanced Grant No. 692670 ``FIRSTORM'' 
and MIUR PRIN 2015 (Prot. 2015C5SEJJ001) and SISSA/CNR project 
"Superconductivity, Ferroelectricity and Magnetism in bad metals" (Prot. 232/2015).
A.V. acknowledges financial support from the Austrian Science Fund
(FWF) through the Erwin Schr\"odinger fellowship J3890-N36.

\bibliography{biblio}

\end{document}